\documentclass[twoside]{article}

\usepackage{PRIMEarxiv}

\usepackage[utf8]{inputenc} 
\usepackage[T1]{fontenc}    
\usepackage{hyperref}       
\usepackage{url}            
\usepackage{booktabs}       
\usepackage{amsfonts}       
\usepackage{nicefrac}       
\usepackage{microtype}      
\usepackage{lipsum}
\usepackage{fancyhdr}       
\usepackage{graphicx}       
\graphicspath{{media/}}     

\usepackage{amsmath}
\usepackage{tikz}
\usepackage{mathdots}
\usepackage{yhmath}
\usepackage{cancel}
\usepackage{color}
\usepackage{siunitx}
\usepackage{array}
\usepackage{multirow}
\usepackage{amssymb}
\usepackage{textcomp}
\usepackage{gensymb}
\usepackage{tabularx}
\usepackage{extarrows}
\usepackage{booktabs}
\usepackage{float}
\usetikzlibrary{fadings}
\usetikzlibrary{patterns}
\usetikzlibrary{shadows.blur}
\usetikzlibrary{shapes}
\usepackage{adjustbox}
\usepackage{comment}

\usepackage{comment}
\usepackage{tikz}
\usetikzlibrary{matrix, calc, fit}
\usetikzlibrary{positioning}
\usepackage{pgfplots}
\graphicspath{{../pdf/}{../jpeg/}}
\DeclareGraphicsExtensions{.pdf,.jpeg,.png}
\usepackage{tabularx}
\usepackage{multirow}
\usepackage{multicol}
\usepackage{lipsum,adjustbox}
\usetikzlibrary{calc,positioning}
\usepackage{pgf-pie}
\usepackage{caption}
\usepackage{subcaption}
\def\checkmark{\tikz\fill[scale=0.4](0,.35) -- (.25,0) -- (1,.7) -- (.25,.15) -- cycle;}
\usepackage{booktabs}
\newcommand{\ignore}[1]{}
\usetikzlibrary{shapes, arrows}
\usepackage{hyperref}
\title{Free-form Shape Modeling in XR: A Systematic Review}

\author{
  Shounak Chatterjee\\
    Kolkata, India\\
{\tt\small shounak.thirti2@gmail.com}
}

\begin{document}

\maketitle
\begin{abstract}
   Shape modeling research in Computer Graphics has been an active area for decades. The ability to create and edit complex 3D shapes has been of key importance in Computer-Aided Design, Animation, Architecture, and Entertainment. With the growing popularity of Virtual and Augmented Reality, new applications and tools have been developed for artistic content creation; real-time interactive shape modeling has become increasingly important for a continuum of virtual and augmented reality environments (eXtended Reality (XR)). Shape modeling in XR opens new possibilities for intuitive design and shape modeling in an accessible way. Artificial Intelligence (AI) approaches generating shape information from text prompts are set to change how artists create and edit 3D models. There has been a substantial body of research on interactive 3D shape modeling. However, there is no recent extensive review of the existing techniques and what AI shape generation means for shape modeling in interactive XR environments. In this state-of-the-art paper, we fill this research gap in the literature by surveying free-form shape modeling work in XR, with a focus on sculpting and 3D sketching, the most intuitive forms of free-form shape modeling. We classify and discuss these works across five dimensions: contribution of the articles, domain setting, interaction tool, auto-completion, and collaborative designing. The paper concludes by discussing the disconnect between interactive 3D sculpting and sketching and how this will likely evolve with the prevalence of AI shape-generation tools in the future.
\end{abstract}


\section{Introduction}
Extended reality, or XR, encompasses all immersive technologies, including Augmented Reality (AR), Virtual Reality (VR), and Mixed Reality (MR). These technologies extend reality by adding to or simulating the natural world through digital materials, blending the virtual and 'real' worlds, or creating a fully immersive experience. In augmented reality, virtual objects and information are overlaid in the real world. The user experiences this through AR glasses or via screens such as a mobile phone or tablet. In virtual reality, users can fully immerse themselves in a simulated digital environment by using a VR headset or head-mounted display to get a 360-degree view of the digital world. In mixed reality, digital and real-world objects co-exist and can interact in real-time. This latest immersive technology is sometimes referred to as hybrid reality.

Shape modeling is a significant aspect of computer graphics, with a rich variety of modeling approaches including 3D shape reconstruction ~\cite{geiger2011stereoscan, izadi2011kinectfusion, mouragnon2006real}, where a user creates or, more precisely, constructs a 3D model from one or multiple RGB or depth images, and statistical shape modeling \cite{davies2001information}, where the generated model is semantically similar to a pre-existing reference. This survey paper focuses on free-form shape modeling, where the user creates and carves a shape as an artistic process. In addition to this, we also analyze the evolution of free-form shape modeling over the past 15 years, particularly through the widespread adoption of virtual reality and augmented reality interaction mechanisms. Devices such as Leap Motion\footnote{https://www.ultraleap.com}, Kinect\footnote{https://azure.microsoft.com/en-us/products/kinect-dk}, HTC Vive\footnote{https://www.vive.com}, Oculus Rift\footnote{https://www.oculus.com}, and Hololens\footnote{https://www.microsoft.com/en-us/hololens} have greatly impacted free-from modeling, especially in virtual sculpting and 3D sketching.

Previously, models were primarily visualized and interacted with on desktop screens or holographic 2D screens \cite{keefeCavePaintingFullyImmersive2001a}. Users could only see a 2D projection of 3D objects and had to rotate the scene to interact with it. Today, mixed reality displays provide designers and artists with a full 3D view and an immersive experience for designing and reviewing 3D shapes intuitively from any viewpoint. Rendering algorithms have also improved over the years, and new visualization techniques enable better modeling by enabling visual insights. In the scope of this survey, we also cover several visualization techniques related to shape modeling.

In this survey paper we give a comprehensive review of the state of the field of free-form shape modeling. Since our research is focused on free-form shape modeling, we discuss  virtual sculpting and 3D sketching as our because these two types of modeling approaches provide users with the freedom to intuitively create models from as an artistic process.



\subsection{Overview}
Although numerous surveys~\cite{olsen2009sketch, dingSurveySketchBased2016, xu2022deep} exist on sketch-based techniques, these focus primarily on elucidating various aspects of both 2D and 3D sketching. We explain these techniques in the following subsection. In our review, we expand on these prior works by also incorporating sculpting techniques to discuss the subject more comprehensively.

\subsubsection{Related Work and Scope}
In 2009, Olsen et al.\cite{olsen2009sketch} published a valuable survey on sketch-based modeling systems. Their work included a systematic categorization, hinging on the interpretation of sketches. Later, in 2016, Ding et al.\cite{dingSurveySketchBased2016} offered their review of sketch-based modeling systems. The authors provided a thorough exposition of different sketch-based modeling methods, including but not limited to 2D sketches, 3D curve sketches, and surface drawings. They classified sketch-based modeling systems based on factors such as input, output, and the modeling approach utilized. More recently, in 2021, Wang et al.~\cite{wang2021ar} conducted a survey on AR and VR remote collaboration. Their work synthesized information on remote collaboration in design, shape reconstruction, and digital sculpting within AR and VR environments. In the following year, 2022, Xu et al.~\cite{xu2022deep} published their research on the state of the art in free-hand sketching within the domain of deep learning. Their study explored the characteristics, challenges, and opportunities related to working with free-hand sketch data. In summary, our survey is positioned to provide a\ignore{broader} combined view, weaving together the insights from the aforementioned research to deliver a comprehensive understanding of both sculpting and 3D sketching techniques.

\subsubsection{Contributions}
Free-form shape modeling which enables users to model any shape by sculpting or sketching. We examine virtual sculpting and 3D sketching in detail and their development in the last 16 years. We followed the guidelines proposed by PRISMA methodology \cite{Pagen160}\footnote{https://www.prisma-statement.org/}  and Kitchenham (2004), and we conducted a systematic literature review. In particular, (a) we analyze how shape modeling methods (section~\ref{sec:shape}) and their user interaction techniques (section~\ref{sec:inter}) evolved in the modern day. (b) we provide a detailed taxonomy of both modeling techniques (section~\ref{sec:shape}), their working environments, and the tools used for interaction (section~\ref{sec:inter}), and (c) we also discuss collaborative shape modeling (section~\ref{sec:collab}) and how artificial intelligence (sections~\ref{sec:AI}) is being used to model and refine a shape. 

\subsubsection{Structure of the review}
The review is organized as follows:
\begin{itemize}
    \item Section ~\ref{sec:method} addresses our methodology that we followed in the manuscript. This section states our research questions. Following that, it gives a rundown of our literature selection process. After submitting our search queries, as described in section ~\ref{sec:method}, we received multiple results that needed to be filtered out since not all of them were pertinent to our study. The selection process is explained in subsection ~\ref{sec:selection}. 
    
    \item In section ~\ref{sec:shape}, we explain the methods of free-form shape modeling. We describe various methods used for sculpting and sketching and how they evolved over time.

    \item In section ~\ref{sec:inter} we describe the different interaction tools. We grouped these into three subsections: gesture-based tool, pen-based tool, and 3D controller. We explain their use and user preference for each type of tool for different sculpting methods.

    \item In sections ~\ref{sec:AI} and ~\ref{sec:collab}, we elucidated the contemporary practices of free-form shape modeling, which encompass collaboration among multiple users and the utilization of artificial intelligence for refining and editing the model. The survey reaches its conclusion in the section~\ref{sec:conclusion}.
    
\end{itemize}


\section{Methodology}\label{sec:method}
Our study provides a systematic literature review of research on free-form shape modeling in mixed reality. The research questions of this study are:

\textbf{RQ1:} What are the most preferred modeling techniques in extended reality?

\textbf{RQ2:} What methodologies/theories are being used to research 3D shape modeling in extended reality?

\textbf{RQ3:} What are the research gaps in shape modeling in extended reality?

\textbf{RQ4:} Which hardware has been preferred for different modeling techniques?

To answer the research questions outlined above, we searched relevant research articles published since 2007 in the following databases and digital libraries: IEEE Xplore, ACM Digital Library, and SpringerLink. We formulated multiple search strings and used primary terms such as 'virtual sculpting,' '3D sketching,' 'free-form shape modeling,' 'mixed reality,' 'augmented reality,' and 'virtual reality.' We also considered alternative spellings and other search terminologies, including related areas, and reviewed the references in the articles. Our search process covered the time frame between October 2021 and October 2023.

\newcommand*{\h}{\hspace{5pt}}
\begin{figure}[!ht]
    \centering
    \begin{tikzpicture}[auto, semithick, remember picture, font=\small,
    block_center/.style ={rectangle, draw=black, thick, fill=white, text width=8em, text centered, minimum height=4em},
    block_left/.style ={rectangle, draw=black, thick, fill=white, text width=10em, text ragged, minimum height=4em},
    block_heading/.style ={rectangle, draw=black, thick, fill=orange!40, text width=24em, text centered, minimum height=2em},
    block_noborder/.style ={rectangle, draw=black, thick, fill=blue!10, text width=22em, text centered, minimum height=1em},
    line/.style ={draw, thick, -latex', shorten >=0pt}]

    \matrix [column sep=5mm,row sep=6mm] {
     \node [block_center] (start) {Record identified from databases (n = 431)}; 
     & \node [block_left] (startR) {Records removed before screening: \\ 
     \h a) Duplicate Records (n=148) \\
     \h b) Records before 2007 (n=59)}; \\
     \node [block_center] (pass1) {Records screened (n=224)}; 
     & \node [block_center] (pass1R) {Records excluded (n=57)};\\
     \node [block_center] (pass2) {Records sought for retrieval (n=167)}; 
     & \node [block_center] (pass2R) {Records not retrieved (n=9)}; \\
     \node [block_center] (pass3) {Records assessed for eligibility (n=158)}; 
     &\node [block_left] (pass3R) {Records excluded: \\ 
     \h Records not related to the scope (n=40)};\\
     \node [block_center] (final) {Studied included for review (n=118)};
	  & \\
    };
    \draw [-stealth](start) -- (startR);
    \draw [-stealth](pass1) -- (pass1R);
    \draw [-stealth](pass2) -- (pass2R);
    \draw [-stealth](pass3) -- (pass3R);
    \draw [-stealth](start) -- (pass1);
    \draw [-stealth](pass1) -- (pass2);
    \draw [-stealth](pass2) -- (pass3);
    \draw [-stealth](pass3) -- (final);

    \coordinate (aux1) at ($(start.west|- startR.north) - (8mm,0mm)$);
    \coordinate (aux2) at ($(start.west|- startR.south) - (1.5mm,0mm)$);
    \node[block_noborder, fit=(aux1)(aux2), inner sep=-.6pt] (X) {}; 
    \node[text width=3cm, text centered, anchor=center, rotate=90] at (X.center) {Identification};

    \coordinate (aux3) at ($(pass1.north west) - (8mm,0mm)$);
    \coordinate (aux4) at ($(pass3.south west) - (1.5mm,0mm)$);
    \node[block_noborder, fit=(aux3)(aux4), inner sep=-.6pt] (Y) {}; 
    \node[text width=3cm, text centered, anchor=center, rotate=90] at (Y.center) {Screening};

    \coordinate (aux5) at ($(final.north west) - (8mm,0mm)$);
    \coordinate (aux6) at ($(final.south west) - (1.5mm,0mm)$);
    \node[block_noborder, fit=(aux5)(aux6), inner sep=-.6pt] (Z) {}; 
    \node[text width=3cm, text centered, anchor=center, rotate=90] at (Z.center) {Included};

    \coordinate (aux7) at ($(X.north west) + (0mm,12mm)$);
    \coordinate (aux8) at ($(startR.north east) + (0mm,3mm)$);
    \node[block_heading, fit=(aux7)(aux8), inner sep=-.6pt] (T) {}; 
    \node[text centered, anchor=center] at (T.center) {Identification of studies via databases and registers};

    \end{tikzpicture}    
    \caption{PRISMA flow diagram for systematic records selection.}
    \label{fig:prismatikz}
\end{figure}
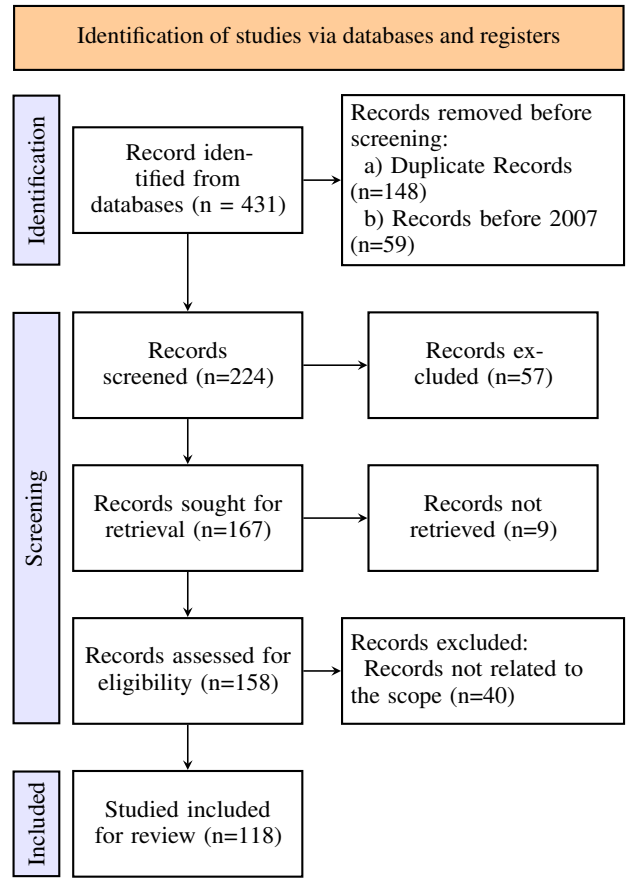

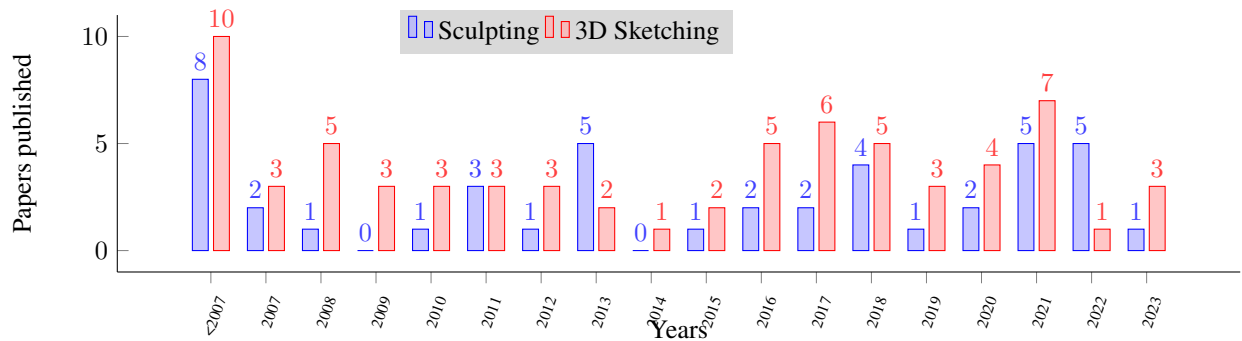
\begin{figure*}[!ht]
\centering
    \begin{tabular}{cc}
     \begin{tikzpicture}
     \begin{axis}[ybar,
    width=\textwidth, height=5 cm,
    enlarge x limits = 0.1,
    legend style={at={(0.4,1.02)},anchor=north,legend columns=-1, fill=gray!30!white, draw=none},
    symbolic x coords = {<2007, 2007, 2008, 2009, 2010, 2011, 2012, 2013, 2014, 2015, 2016, 2017, 2018, 2019, 2020, 2021, 2022, 2023},
    axis x line*=bottom,
    axis y line*=left,
    bar width=6pt,
    ylabel={Papers published},
    xlabel={Years},
    xtick=data,
    nodes near coords,
    nodes near coords align={auto},
    every axis plot/.append style={fill,fill opacity=0.75},
    x tick label style = {font = \tiny, align = center, rotate = 70, anchor = north east},]
    \addplot coordinates{(<2007,8) (2007,2) (2008,1) (2009,0) (2010,1) (2011,3) (2012,1) (2013,5) (2014,0) (2015,1) (2016,2) (2017,2) (2018,4) (2019,1) (2020,2) (2021,5) (2022,5) (2023,1)};
    \addplot coordinates{(<2007,10) (2007,3) (2008,5) (2009,3) (2010,3) (2011,3) (2012,3) (2013,2) (2014,1) (2015,2) (2016,5) (2017,6) (2018,5) (2019,3) (2020,4) (2021,7) (2022,1) (2023,3)};
     \legend{Sculpting, 3D Sketching}
     \end{axis}
     \end{tikzpicture} \\
\end{tabular}
\caption{Number of papers published over the years on Sculpting and 3D sketching.}
\label{trend}
\end{figure*}

\begin{figure}[!ht]
    \centering
    \begin{tikzpicture}
     \tikzset{lines/.style={draw=none},}
    \tikzstyle{every node}=[font=\tiny]
     \pie[ style={lines}, sum=auto,]
     {
     3/3DUI,
     16/CHI,
     4/I3D,
     6/IEEEVR,
     4/ISMAR,
     9/SIGGRAPH,
     4/TVCG,
     11/UIST,
     4/VRST,
     14/TOG,
     4/EG,
     14/CVPR,
     5/ICCV
     19/Others
     }
    \end{tikzpicture}    
    \caption{Number of publications according to the publication venue in our survey duration.}
    \label{fig:venue}
\end{figure}
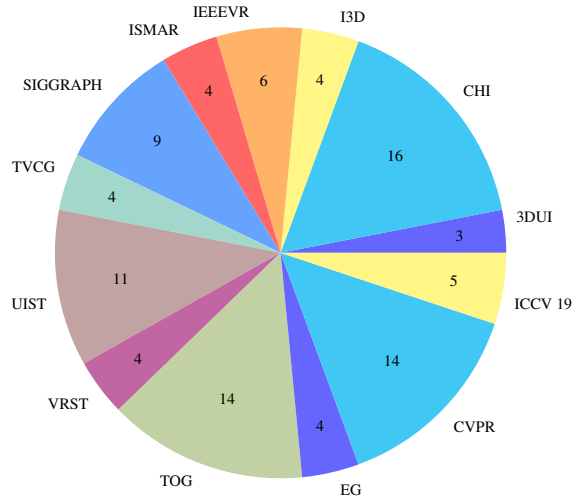

\subsection{Selection Process}\label{sec:selection}
After performing the search queries, we reviewed the publications and applied inclusion and exclusion on the following criteria. We included conference proceedings and journal articles published from 2007 onward and assessed the relevance of each paper based on its title and abstract. We only considered papers related to free-form shape modeling in extended reality environments (mixed, virtual, or augmented) or  2D screens (desktop, drawing tablets, etc.).

Further, we critically appraised the research papers and grouped these based on the type of modeling techniques, the types of environments where the modeling was performed, and the interaction tools used for modeling. Out of 431 papers retrieved using our provided keywords, 224 articles passed the first stage, and 167 contained relevant studies. Finally, we selected 109 papers as primary sources for our review. In addition to these papers, we also included some studies (around 15\% of the total papers) published before 2007 that provided an underpinning foundational explanation behind the current studies. In Figures ~\ref{trend} and ~\ref{fig:venue} we have shown the statistics on the research selected for our review, and in Figure ~\ref{fig:prismatikz} we have shown the article selection process in a flow diagram. In Table ~\ref{tab:sculpting} and ~\ref{tab:sketching} we have listed a few selected papers from our study based on their popularity, i.e. the number of citations and publishing venue, and impact, i.e., how their method is used in the future research work repeatedly.

\subsection{Types of literature contribution}\label{sec:contri}
In the first part of our study, we conducted a literature review based on the shape modeling techniques. Free-from shape modeling can be categorized into two primary techniques: sculpting and sketching. In sculpting, a user starts from a primitive shape and carves out a desired shape. In contrast, sketching involves starting from scratch to create a set of surface patches from 3D strokes, which eventually lead to a shape. Irrespective of the specific modeling techniques employed, the majority of the reviewed papers focused on either introducing novel algorithms for shape representation and modeling, proposing innovative interaction methods, or occasionally even both. In this review, we have grouped and described each modeling technique according to its free-form modeling type, mentioned in section~\ref{sec:shape}. Additionally, we discussed the involvement of the interactive tools, which we classified into three categories and explained their use and preference for particular modeling techniques in section~\ref{sec:inter}.

\begin{center}
    \begin{table*}[!htp]
    \centering
    \caption{Categorization of different sculpting techniques published. These papers are sorted year-wise.}
    \label{tab:sculpting}
     \begin{tabular}{@{}lcccclccc@{}}
        \toprule
        \multirow{2}{*}{Methods} & \multirow{2}{*}{Sculpting Method} & \multicolumn{3}{c}{Environment} &  & \multicolumn{3}{c}{Interaction Tool} \\ \cmidrule(l){3-9} 
         &  & Desktop & AR & VR &  & Hand-Gesture & Controller & Stylus \\ \midrule
        Sculpting \cite{galyeanSculptingInteractiveVolumetric1991a} & Virtual Clay & \checkmark &  &  &  &  & \checkmark &  \\
        NURBS \cite{lamousinNURBSbasedFreeformDeformations1994} & Surface Modeling & \checkmark &  &  &  &  &  &  \\
        Virtual 3D Sculpting \cite{wongVirtual3DSculpting2000a} & Surface Modeling & \checkmark &  & \checkmark &  & \checkmark &  &  \\
        Surface Drawing \cite{schkolneSurfaceDrawingCreating2001a} & Surface Modeling &  &  & \checkmark &  & \checkmark &  & \checkmark \\
        Virtual Clay \cite{mcdonnellVirtualClayRealtime2001a} & Virtual Clay & \checkmark &  &  &  & \checkmark &  &  \\
        6DOF \cite{liveraniEfficient6DOFTools2004} & Surface Modeling &  &  &  &  &  &  &  \\
        V3DSPP \cite{shengInterfaceVirtual3D2006} & Physical Proxy & \checkmark &  &  &  & \checkmark &  &  \\
        ARPottery \cite{hanARPotteryExperiencing2007} & Virtual Clay &  & \checkmark &  &  & \checkmark &  &  \\
        Hands on Virtual Clay \cite{pihuitHandsVirtualClay2008} & Virtual Clay & \checkmark &  &  &  & \checkmark &  &  \\
        AFSC3DM \cite{marnerAugmentedFoamSculpting2010a} & Physical Proxy & \checkmark & \checkmark &  &  &  & \checkmark &  \\
        deForm \cite{follmerDeFormInteractiveMalleable2011} & Physical Proxy &  &  &  &  & \checkmark &  & \checkmark \\
        FreeD \cite{zoranFreeDFreehandDigital2013b} & Physical Proxy & \checkmark &  &  &  &  &  &  \\
        SMNs \cite{stanculescuSculptingMultidimensionalNested2013e} & Multi-layer &  &  &  &  &  &  & \checkmark \\
        MEMs \cite{milliezMutableElasticModels2013b} & Virtual Clay & \checkmark &  &  &  &  &  &  \\
        CSculpt \cite{calabreseCSculptSystemCollaborative2016b} & AI Sculpting &  &  &  &  &  &  &  \\
        SLayer \cite{calabreseSLayerSystemMultiLayered2017b} & Multi-layer &  &  &  &  &  &  &  \\
        Auto-complete 3D Sculpting \cite{pengAutocomplete3DSculpting2018a} & AI Sculpting &  &  &  &  &  &  &  \\
        DigiClay \cite{gaoDigiClayInteractiveInstallation2018a} & Virtual Clay &  & \checkmark &  &  &  &  &  \\
        RealPot \cite{gaoRealPotImmersiveVirtual2019} & Virtual Clay &  &  & \checkmark &  &  & \checkmark &  \\
        Mesh R-CNN \cite{gkioxari2019mesh} & AI Sculpting & \checkmark &  &  &  & \checkmark &  &  \\
        DeepSDF \cite{park2019deepsdf} & AI Sculpting & \checkmark &  &  &  & \checkmark &  &  \\
        DARLAT \cite{natakuaithungDevelopmentARLearning2020a} & Physical Proxy & \checkmark &  &  &  &  &  & \checkmark \\
        PotteryVR \cite{dashtiPotteryVRVirtualReality2022} & Virtual Clay &  &  & \checkmark &  &  & \checkmark &  \\
        DreamFusion \cite{poole2022dreamfusion} & AI Sculpting & \checkmark &  &  &  &  &  &  \\
        NerfShop \cite{jambon2023nerfshop} & AI Sculpting & \checkmark &  &  &  &  &  &  \\ \bottomrule
        \end{tabular}%
    \end{table*}
\end{center}

\section{Free-form Shape modeling}\label{sec:shape}
Research in computer graphics has produced various essential and powerful techniques for image and shape synthesis. While shape modeling and processing techniques are well understood, user-interface methods for creating and manipulating computer models of 3D shapes are still relatively difficult in the desktop environment than modeling in mixed reality space~\cite{kwanMobi3DSketch3DSketching2019, aroraExperimentalEvaluationSketching2017}. Consequently, 3D shape modeling is a very laborious procedure that requires considerable human time and effort in a 2D visualization. In our literature review, we found that previous research indicates that users typically prefer to perform modeling in a 3D space where they have enough freedom to make use of the extra dimension. This extra dimension is helpful for modeling purposes and is also an excellent way to demonstrate and review the model with others~\cite{keefeScientificSketchingCollaborative2008, jacksonLiftOffUsingReference2016}.

\subsection{Sculpting}\label{sec:sculpt}
Sculpting is one of the key free-form shape modeling techniques. Here, a user starts with a pre-existing basic shape of a 3D object. The user can perform various operations such as extruding, indenting, and carving on the surface of an object to give a shape of choice. Usually, users shape the object by hand or with a special tool. Certain sculpting literature delves into various aspects, such as algorithms, elucidating different approaches to representing shapes. Another facet explored in the literature involves exploring interaction tools and techniques, including traditional and unconventional instruments employed to engage with the objects, sometimes within alternative realities or environments. In our study, we found different types of sculpting techniques that are explained below. In Table~\ref{tab:sculpting}, we have chronologically listed some standout studies from our review based on their impact metric explained earlier. 

Digital or virtual sculpting was introduced back in 1991 by Galyean et al. \cite{galyeanSculptingInteractiveVolumetric1991a}. The authors represented the primitive shape, i.e., the shape from which the user started sculpting, in a mesh data structure and recorded the changes made by the interaction tool (3D controller) in a voxel grid space. The voxel data is converted to a mesh using the Marching Cubes algorithm~\cite{lorensen1998marching}. In contrast to this, Wong et al. \cite{wongVirtual3DSculpting2000a} proposed a method to directly control the shape using a control hand surface to fit open uniform B-spline surfaces. Since the authors used an electronic glove as their input device, they took hand points (joints and fingertips) to create a control surface. This control surface directed the sculpting operations. A set of parametric surfaces represented a shape, and it was updated by interpolating the changes in the hand points received from the user.

\subsubsection{Virtual Clay}
\begin{figure}[!htp]
    \centering
    \begin{subfigure}{.45\textwidth}
     \centering
     \includegraphics[width=.95\linewidth]{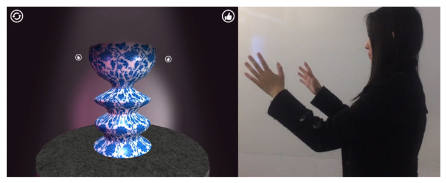}  
     \caption{DigiClay \cite{gaoDigiClayInteractiveInstallation2018a}}
     \label{SUBFIGURE LABEL 1}
    \end{subfigure}
    \begin{subfigure}{.45\textwidth}
     \centering
     \includegraphics[width=.95\linewidth]{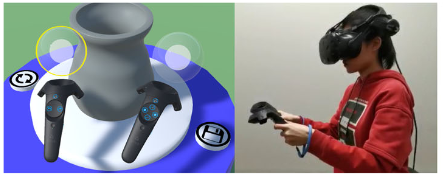}  
     \caption{RealPot \cite{gaoRealPotImmersiveVirtual2019}}
     \label{SUBFIGURE LABEL 2}
    \end{subfigure}
    \begin{subfigure}{.45\textwidth}
     \centering
     \includegraphics[width=.95\linewidth]{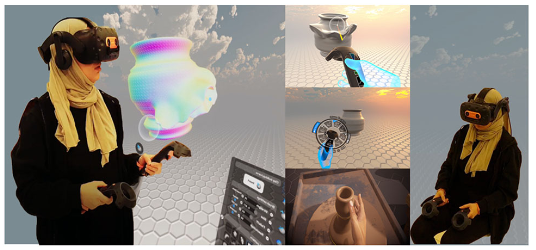}  
     \caption{PotteryVR \cite{dashtiPotteryVRVirtualReality2022}}
     \label{SUBFIGURE LABEL 3}
    \end{subfigure}
    \caption{Different techniques for VR sculpting. (a) Sculpting with hand gestures; (b) \& (c) sculpting the model using VR controllers.}
    \label{fig:foobar}
\end{figure}

McDonnell et al. \cite{mcdonnellVirtualClayRealtime2001a} had a different approach to sculpting, where the starting object was represented as a virtual 3D elastic clay. Virtual clay is a digital sculpting technique that mimics the experience of working with physical clay. Through this method, users can shape virtual objects using tools like a virtual sculpting brush or input devices such as a 3D mouse or a stylus. The authors propose a dynamic subdivision solid modeling approach, where the object's surface is dynamically divided into smaller polygons. This allows users to sculpt specific portions of the object, resulting in localized changes rather than altering the entire shape. In the sculpting workspace, the virtual clay reacts to the user's direct application of forces in an intuitive and predictable way, and the haptic feedback can greatly improve the sense of realism. Many researchers further adopted this virtual clay system \cite{hanARPotteryExperiencing2007, pihuitHandsVirtualClay2008, gaoDigiClayInteractiveInstallation2018a, gaoRealPotImmersiveVirtual2019, barreiroNaturalTactileInteraction2021, dashtiPotteryVRVirtualReality2022} to account for different environments and tools. Han et al.~\cite{hanARPotteryExperiencing2007}  designed an AR-based pottery system based on the virtual clay system. The users can make the pottery intuitive and use one or both hands to perform different interaction functions like poke, push, pull, bend, stretch, etc. Pihuit et al.~\cite{pihuitHandsVirtualClay2008} proposed a system with a passive tactile feedback device to shape the virtual clay with hands. It allows users to apply finger pressure on the pressure sensors, and it returns tactile feedback and realistic visual behavior of the hand to provide a better immersion for the user. Milliez et al. \cite{milliezMutableElasticModels2013b} and Gao et al. \cite{gaoDigiClayInteractiveInstallation2018a} proposed a similar virtual clay representation. In the former, authors proposed to add virtual elastic clay models to create a final shape instead of using a surface. The goal was to add separate models and blend them. They incorporated automatic local adaptation into the sculpting operation to accommodate model updates and deformations. The latter used real-time mesh deformation of the virtual clay using hands to provide users with an effortless and refined interface for art creation. Gao et al. \cite{gaoRealPotImmersiveVirtual2019} proposed a virtual pottery system that is compatible with VR. They used a cylinder-shaped clay mesh as their starting point and gave it a shape by deforming the mesh and changing the radius at different heights. Users have the option to update the height, thickness, and non-uniform radius. In Figure ~\ref{fig:foobar}, we have shown different sculpting techniques in VR environment adapted from ~\cite{gaoDigiClayInteractiveInstallation2018a, gaoRealPotImmersiveVirtual2019, dashtiPotteryVRVirtualReality2022}.

Later, Dashti et al.~\cite{dashtiPotteryVRVirtualReality2022} presented a scalable virtual pottery approach named PotteryVR, which can create more realistic forms in the virtual world. Their system used commercially available VR equipment, which is more accessible to the users and it can accommodate multiple users. The authors also discussed the evaluation methods for VR, focusing on visual feedback, gestural interfaces, 3D modeling, and collaborative design. Their study aims to develop modeling and practice skills, provide guidelines on designing 3D models, and use of visual pottery for deformable shape modeling and rapid prototyping. Barrerio et al.~\cite{barreiroNaturalTactileInteraction2021} presented another virtual clay modeling system with natural tactile interaction. Their solution involved utilizing a clay simulation method and an ultrasound-based rendering algorithm. This system effectively handles extreme viscoplasticity and offers superior interaction capabilities compared to previous methods. However, it has limitations, including the high computational cost of the elastoplasticity constraint, restricted rendering method coverage, and limited detailed finger modeling dexterity due to low particle resolution and limited tactile stimulation( see Figure~\ref{fig:foobar}).

\subsubsection{Surface Modeling}
In addition to the virtual clay system, another prevalent method involves users modifying a hollow object's surface. Lamousin et al.~\cite{lamousinNURBSbasedFreeformDeformations1994} proposed a surface modeling method utilizing non-uniform rational B-splines (NURBS) to modify curves and surfaces in 3D modeling. 
Unlike virtual clay, the NURBS-based surface modeling method only modifies the object's surface. In this modeling approach, the objects utilized are typically represented using implicit curves or surfaces~\cite{lamousinNURBSbasedFreeformDeformations1994}. Sometimes, the implicit surfaces are discretized into mesh data structures~\cite{liveraniEfficient6DOFTools2004}; sometimes, the curves are tessellated with mesh as well. A set of control points defines the shape of the surface, and a set of weights controls the influence of each control point on the shape. Research by Liverani et al.~\cite{liveraniEfficient6DOFTools2004} and Schkolne et al.~\cite{schkolneSurfaceDrawingCreating2001a} used NURBS in different environments and interaction tools.

\subsubsection{Physical Proxy}
Sheng et al.~\cite{shengInterfaceVirtual3D2006} also showed mesh manipulation in their work by using NURBS with a physical prop as a proxy to the virtual model for better control of the interaction techniques. Later, researchers adopted the proxy method in various ways. 
Marner et al.~\cite{marnerAugmentedFoamSculpting2010a} proposed to physically sculpt a 3D model from foam using traditional tools. In that process, the system digitally duplicates the process by tracking the tools and the foam. A similar method was adopted by Zoran et al.~\cite{zoranFreeDFreehandDigital2013b} with a better foam sculptor. There are many similar methods for converting physical sculpting to virtual models~\cite{hattab3DModelingScanning2015, ohPEP3DPrinted2018a}. Follmer et al.~\cite{follmerDeFormInteractiveMalleable2011} and Natakuaithung et el.~\cite{natakuaithungDevelopmentARLearning2020a} proposed a similar approach but used modeling clay instead of foam. 

\subsubsection{Multi-layer Sculpting}
All the sculpting methods explained before had only one object, in either virtual clay or mesh data structure. However, a model can be constructed with multiple sub-models~\cite{stanculescuSculptingMultidimensionalNested2013e, milliezMutableElasticModels2013b}, or it can be sculpted with multiple layered materials~\cite{calabreseSLayerSystemMultiLayered2017b}. Schmidt et al.~\cite{schmidtAnalyticDrawing3D2009} proposed the idea of fusing two meshes by \textit{drag-and-drop} methods. Their approach was to add another mesh by filling the holes of a prior mesh. Stanculescun et al.~\cite{stanculescuSculptingMultidimensionalNested2013e} presented a systematic approach to representing and deforming shapes with nested features to handle free-form deformations. They used a hierarchical rule-based approach with separate attributes for shape and geometry to adjust deformations in a sculpting system for mesh features like regions, curves, and points. Milliez et al.~\cite{milliezMutableElasticModels2013b} introduced a deformation model for sculpting shapes with local adjustments and geometry changes. This approach made the model more flexible than traditional methods. The smart-clay application took advantage of this flexibility, adjusting shapes through methods like stretching, merging, and splitting in response to desired changes.

\subsection{3D Sketching}\label{sec:sketch}
3D sketching is another type of popular free-form shape modeling. To create a surface or an object, the user scribbles some strokes in space, and those become a collection of curves or surfaces. The user either starts from scratch or adds something on top of a pre-existing shape without deforming it. Similar to sculpting, different sketching techniques exist, especially depending on the interactive tools. In Table~\ref{tab:sketching}, we have listed some of this research on sketching, chronologically based on\ignore{impact on this field} their popularity, i.e. the number of citations and publishing venue, and impact, i.e., how their method is used in the future research as we mentioned previously.

During the 1990s and early 2000s, 3D sketching was performed to create 3D geometric shapes, which were manipulated using editing tools. The overall process resembled sculpting, with initial ideas borrowed from sculpting techniques~\cite{deeringHoloSketchVirtualReality1995, keefeCavePaintingFullyImmersive2001a, wongVirtual3DSculpting2000a, galyeanSculptingInteractiveVolumetric1991a}. Deering et al.~\cite{deeringHoloSketchVirtualReality1995} introduced a 2D sketch paradigm that could be expanded to 3D in VR, supporting complex animated 3D scenes. However, at that time, drawing a 3D curve required users to examine the projection of the curve from different viewpoints on a monitor~\cite{cohen1999interface}. Cohen et al. ~\cite{cohen1999interface} proposed a novel method for specifying 3D curves using 2D input from a single viewpoint, utilizing the shadow of the curve on the floor. Additionally, Igarashi et al.~\cite{igarashiTeddySketchingInterface2006a} conducted research on creating smooth 3D shapes from 2D free-form sketches.

\subsubsection{Generating 3D shapes from 2D sketches}
There have been numerous research studies following the work of Igarashi et al. ~\cite{igarashiTeddySketchingInterface2006a} that utilized techniques to generate 3D shapes from 2D sketches, representing the resulting shapes meshes ~\cite{igarashiTeddySketchingInterface2006a, wangFreeformSketch2007, eitzSketchbasedShapeRetrieval2012, ramosNewUserFriendlySketchBased2016, guillardSketch2MeshReconstructingEditing2021a, xuTrue2Form3DCurve2014, karpenkoSmoothSketch3DFreeform2006}, voxels~\cite{kaziDreamSketchEarlyStage2017, zhangSketch2ModelViewAware3D2021}, and parametric surfaces~\cite{chenSketchingRealityRealistic2008a, liu3DReconstructionFreeform2010, ye3DSketchbased3D2016}.

The research conducted by Igarashi et al.\cite{igarashiTeddySketchingInterface2006a} proposed a system that takes a 2D sketch as input and inflates it into a 3D object. Additional details can be added to the object by scribbling on it. Wang et al.\cite{wangFreeformSketch2007} proposed modeling complex 3D shapes by inflating 2D sketches, using layered meshes to combine primitive shapes into a complex 3D structure. Gingold et al.~\cite{gingoldStructuredAnnotations2Dto3D2009} developed a similar system for generating 3D models from 2D sketches, but with the requirement for users to annotate and position the primitive shapes within the sketch to provide semantic information, aiming to improve representation. Their goal was to eliminate the constant need for rotating the sketch/model during the sketching and modification process. Liu et al.\cite{liu3DReconstructionFreeform2010} proposed a method for constructing 3D free-form surface patches from 2D line drawings. Their approach included techniques for curve interpolation from the boundary, reconstruction of 3D curves from their 2D projections, and reconstruction of surfaces from their backbones, which are curves inside the input boundary sketch. Chen et al.\cite{chenSketchingRealityRealistic2008a} introduced a process for converting sketches into a realistic 2.5D interpretation of an architectural design sketch with editing options.

Ramos et al.~\cite{ramosNewUserFriendlySketchBased2016} proposed a sketch-based modeling system that uses convolution surfaces. The user draws their conceptual ideas in 2D planes embedded in 3D space. The 2D drawing is then projected onto 3D surfaces, and the depth in space is computed via ray-casting. Regarding 2D sketches, Xu et al.\cite{xuTrue2Form3DCurve2014} used 2D vector sketches to generate 3D curves and employed mesh structures to generate a closed object. Guillard et al.\cite{guillardSketch2MeshReconstructingEditing2021a} and Zhang et al.~\cite{zhangSketch2ModelViewAware3D2021} used 2D sketches and utilized neural networks to generate 3D objects in mesh and voxel structures, respectively.

\begin{table*}[!htp]
    \centering
    \caption{Categorization of different sketching techniques published. These papers are sorted year-wise.}
    \label{tab:sketching}
    \resizebox{\textwidth}{!}{%
    \begin{tabular}{@{}lcccclccc@{}}
        \toprule
        \multirow{2}{*}{Method} & \multirow{2}{*}{Sketching Technique} & \multicolumn{3}{c}{Envirionment} &  & \multicolumn{3}{c}{Interaction Tool} \\ \cmidrule(l){3-9} 
         &  & Desktop & AR & VR &  & Hand-Gesture & Controller & Stylus \\ 
         \midrule
        An interface for sketching 3d curves \cite{cohen1999interface} & 3D Curve Sketching & \checkmark &  &  &  &  &  &  \\
        CavePainting \cite{keefeCavePaintingFullyImmersive2001a} & 3D Curve Sketching &  &  & \checkmark &  &  &  & \checkmark \\
        FreeDrawer \cite{wescheFreeDrawerFreeformSketching2001} & 3D Curve Sketching & \checkmark &  &  &  &  &  & \checkmark \\
        Teddy \cite{igarashiTeddySketchingInterface2006a} & 2D-to-3D Sketching &  & \checkmark &  &  &  & \checkmark & \checkmark \\
        Free-form Sketch \cite{wangFreeformSketch2007} & 2D-to-3D Sketching & \checkmark &  &  &  &  &  & \checkmark \\
        Drawing in Air \cite{keefeDrawingAirInput2007a} & 3D Curve Sketching &  &  & \checkmark &  & \checkmark &  & \checkmark \\
        FiberMesh \cite{nealenFiberMeshDesigningFreeform2007} & 3D Sketch-to-Shape & \checkmark &  &  &  &  &  & \checkmark \\
        Sketch3D \cite{fang2008interactive} & 3D Curve Sketching &  &  &  &  &  &  & \checkmark \\
        ILoveSketch \cite{baeILoveSketchAsnaturalaspossibleSketching2008a} & 3D Curve Sketching & \checkmark &  &  &  &  &  & \checkmark \\
        Sketching Reality \cite{chenSketchingRealityRealistic2008a} & 2D-to-3D Sketching & \checkmark &  &  &  &  &  & \checkmark \\
        Napkin sketch\cite{xinNapkinSketchHandheld2008a} & 3D Curve Sketching &  & \checkmark &  &  &  &  & \checkmark \\
        Structured annotations for 2D-to-3D modeling \cite{gingoldStructuredAnnotations2Dto3D2009} & 2D-to-3D Sketching & \checkmark &  &  &  &  &  & \checkmark \\
        Analytic drawing of 3D scaffolds \cite{schmidtAnalyticDrawing3D2009} & 3D Curve Sketching & \checkmark &  &  &  &  &  & \checkmark \\
        EverybodyLovesSketch \cite{baeEverybodyLovesSketch3DSketching2009a} & 3D Curve Sketching & \checkmark &  &  &  &  &  & \checkmark \\
        3DCurveSketch \cite{fabbri3DCurveSketch2010} & 3D Curve Sketching & \checkmark &  &  &  &  &  & \checkmark \\
        OverCoat \cite{schmidOverCoatImplicitCanvas2011} & 3D Curve Sketching & \checkmark &  &  &  &  &  & \checkmark \\
        Single-view sketch based modeling \cite{andreSingleviewSketchBased2011} & 3D Sketch-to-Shape & \checkmark &  &  &  &  &  & \checkmark \\
        JustDrawIt \cite{grimmJustDrawIt2012} & 3D Curve Sketching & \checkmark &  &  &  &  &  &  \\
        Sketch-based shape retrieval \cite{eitzSketchbasedShapeRetrieval2012} & 2D-to-3D Sketching & \checkmark &  &  &  &  &  & \checkmark \\
        SecondSkin \cite{depaoliSecondSkinSketchbasedConstruction2015} & 3D Curve Sketching &  &  &  &  &  &  & \checkmark \\
        3D Sketch-Based 3D Model Retrieval \cite{li3DSketchBased3D2015} & 3D Sketch-to-Shape & \checkmark & \checkmark &  &  &  &  & \checkmark \\
        SmartCanvas \cite{zhengSmartCanvasContextinferredInterpretation2016} & Model Guided Sketching &  &  &  &  &  &  & \checkmark \\
        Lift-Off \cite{jacksonLiftOffUsingReference2016} & Model Guided Sketching &  &  & \checkmark &  &  &  & \checkmark \\
        3D Shape Reconstruction from Sketches via MVCN \cite{lun3DShapeReconstruction2017a} & 3D Sketch-to-Shape &  &  &  &  &  &  & \checkmark \\
        WireDraw \cite{yueWireDraw3DWire2017a} & Model Guided Sketching &  &  & \checkmark &  &  &  & \checkmark \\
        DreamSketch \cite{kaziDreamSketchEarlyStage2017} & 2D-to-3D Sketching &  &  & \checkmark &  &  &  & \checkmark \\
        SweepCanvas \cite{liSweepCanvasSketchbased3D2017} & 3D Curve Sketching &  & \checkmark &  &  &  &  &  \\
        EESSVR \cite{aroraExperimentalEvaluationSketching2017} & 3D Curve Sketching &  &  & \checkmark &  &  &  & \checkmark \\
        SymbosisSketch \cite{aroraSymbiosisSketchCombining2D2018} & 3D Curve Sketching &  &  &  &  &  &  & \checkmark \\
        Model-Guided 3D Sketching \cite{xuModelGuided3DSketching2019} & Model Guided Sketching &  & \checkmark &  &  &  &  & \checkmark \\
        VRSketchPen \cite{elsayedVRSketchPenUnconstrainedHaptic2020} & 3D Curve Sketching &  &  & \checkmark &  &  & \checkmark & \checkmark \\
        VRSketchIn \cite{dreyVRSketchInExploringDesign2020} & 3D Curve Sketching &  &  & \checkmark &  &  &  & \checkmark \\
        HandPainter \cite{jiangHandPainter3DSketching2021} & 3D Curve Sketching &  &  & \checkmark &  &  &  & \checkmark \\
        Sketch2Mesh \cite{guillardSketch2MeshReconstructingEditing2021a} & 2D-to-3D Sketching & \checkmark &  &  &  &  &  & \checkmark \\
        Sketch2Model \cite{zhangSketch2ModelViewAware3D2021} & 2D-to-3D Sketching & \checkmark &  &  &  &  &  & \checkmark \\
        CASSIE \cite{yuCASSIECurveSurface2021} & 3D Curve Sketching &  &  & \checkmark &  &  & \checkmark &  \\
        ScaffoldSketch \cite{yuScaffoldSketchAccurateIndustrial2021} & 3D Curve Sketching &  &  & \checkmark &  &  &  & \checkmark \\
        Mid-Air Drawing of Curves on 3D Surfaces in VR \cite{aroraMidAirDrawingCurves2021a} & 3D Curve Sketching &  &  & \checkmark &  &  & \checkmark &  \\
        PSFU3DS \cite{yuPiecewisesmoothSurfaceFitting2022} & 3D Curve Sketching &  &  & \checkmark &  &  & \checkmark &  \\
        SKED \cite{mikaeili2023sked} & AI Sketching &  &  &  &  &  &  & \checkmark \\ 
        \bottomrule
    \end{tabular}
    }
\end{table*}

\subsubsection{3D curve Sketching}
\begin{figure}[!ht]
    \centering
    \begin{subfigure}{.45\textwidth}
     \centering
     \includegraphics[width=.95\linewidth]{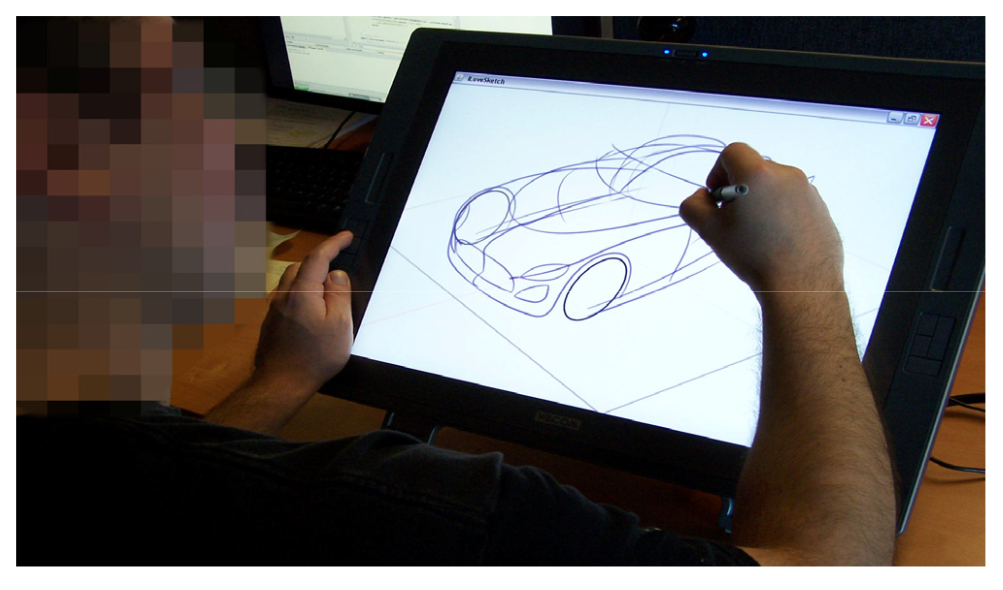}  
     \caption{ILoveSketch \cite{baeILoveSketchAsnaturalaspossibleSketching2008a}}
     \label{SUBFIGURE LABEL 4}
    \end{subfigure}
    \begin{subfigure}{.45\textwidth}
     \centering
     \includegraphics[width=.95\linewidth]{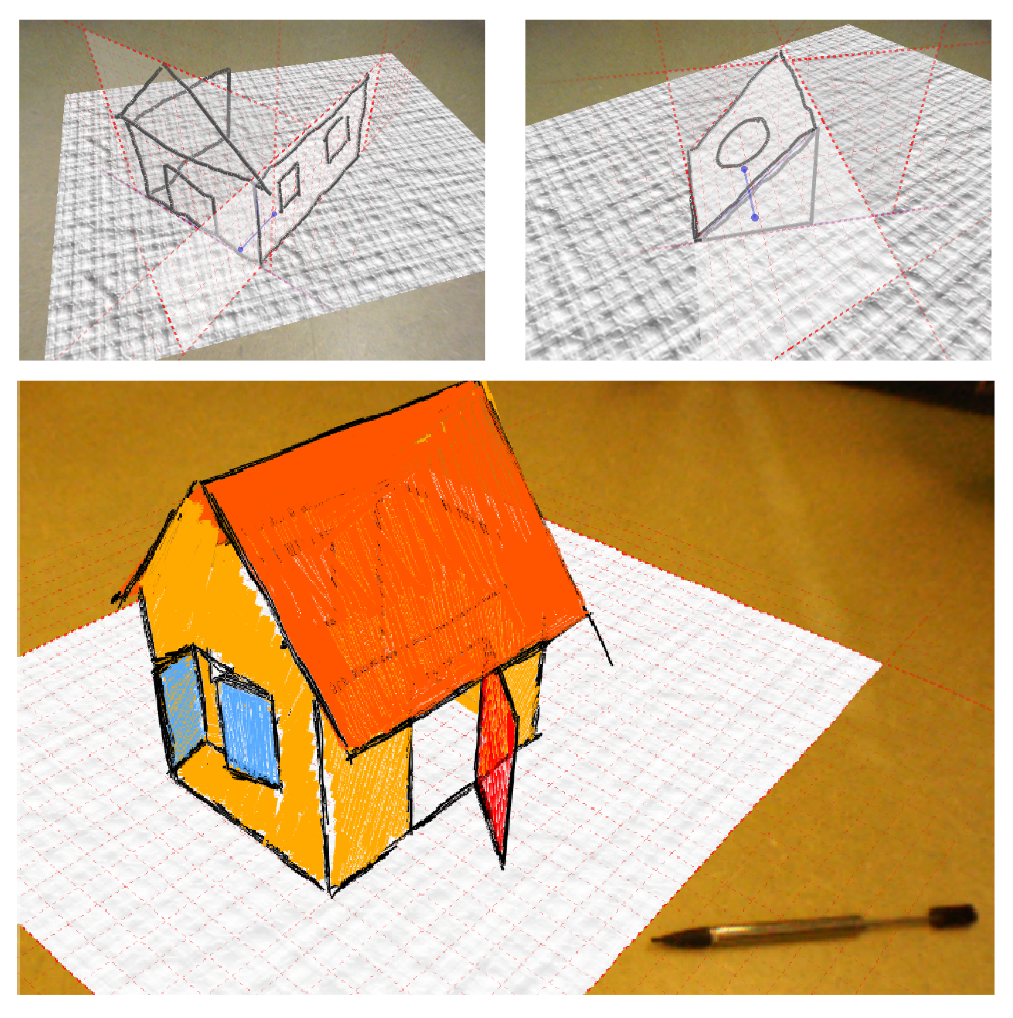}  
     \caption{Napkin Sketch \cite{xinNapkinSketchHandheld2008a}}
     \label{SUBFIGURE LABEL 5}
    \end{subfigure}
    \begin{subfigure}{.45\textwidth}
     \centering
     \includegraphics[width=.95\linewidth]{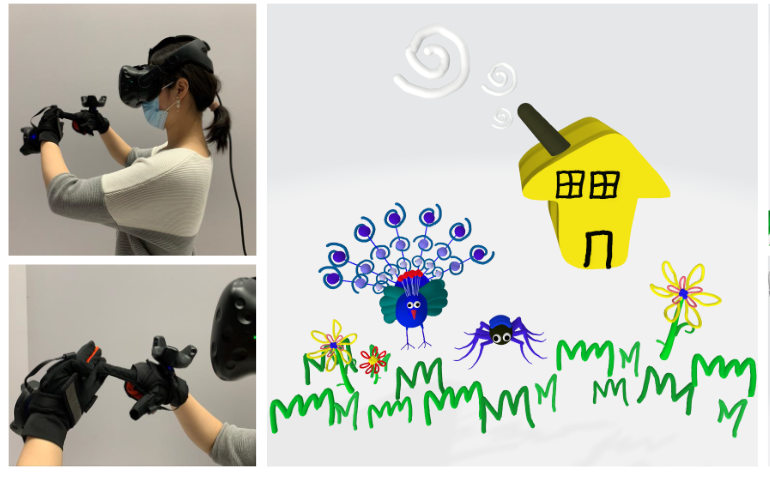}  
     \caption{HandPainter \cite{jiangHandPainter3DSketching2021}}
     \label{SUBFIGURE LABEL 6}
    \end{subfigure}
    \caption{Different techniques of 3D sketching. (a) Sketching on top of a drawing tablet screen; (b) Sketching in AR on a designated area; (c) Drawing in VR.}
    \label{fig:foobar2}
\end{figure}

In this subsequent subsection, we explain different types of 3D sketching. During our study on 3D sketching, we observed that these methods often involve creating 3D sketches and sometimes enhancing the curves with virtual surfaces or generating surfaces within a 3D space. Moreover, there are instances where models are retrieved based on the 3D sketches provided by the users. We have categorized the methods according to these specific techniques. There are a variety of 3D curve sketching techniques. In some methods, the user designs an object using 3D curves or surfaces, which may subsequently be tessellated with meshes~\cite{schkolneSurfaceDrawingCreating2001a, baeILoveSketchAsnaturalaspossibleSketching2008a, yuCASSIECurveSurface2021, xinNapkinSketchHandheld2008a, jiangHandPainter3DSketching2021}. In other technique, users construct a scene on a 3D canvas employing both curves and surfaces~\cite{keefeCavePaintingFullyImmersive2001a, liSweepCanvasSketchbased3D2017, aroraSymbiosisSketchCombining2D2018, dreyVRSketchInExploringDesign2020}. In Figure~\ref{fig:foobar2}, we compare some of the sketching methods in different environments adopted from the methods~\cite{baeILoveSketchAsnaturalaspossibleSketching2008a, xinNapkinSketchHandheld2008a, jiangHandPainter3DSketching2021}.

In the year 1999, Cohen et al.~\cite{cohen1999interface} described a process to draw 3D curves using a single 2D viewpoint. After that, we found several research works on 3D sketching. As an early research, Wesche et al. ~\cite{wescheFreeDrawerFreeformSketching2001} proposed a 3D sketching workbench consisting of 3D tools for curve sketching and surface drawing. The system was designed in a virtual environment and uses a stylus as the input method. Another remarkable system was designed by Schkolne et al.~\cite{schkolneSurfaceDrawingCreating2001a}, where users can design 3D shapes from the surface marked by their hands in a 3D virtual space. Keefe et al.~\cite{keefeDrawingAirInput2007a} introduced haptic-aided drawing of 3D curves in a virtual space. They demonstrated that for drawing in 3D space, haptic feedback is necessary for precise control and accuracy. Their user study showed that users of their system were able to draw complex shapes. Bae et al.~\cite{baeILoveSketchAsnaturalaspossibleSketching2008a, baeEverybodyLovesSketch3DSketching2009a} proposed a sketching technique on a 2D screen, i.e., drawing tablet. Their system was in a tablet-stylus setting, so it could replicate the pen-paper feeling to the professional designers. The authors maintain the accuracy of the 3D curves by projecting objects in multiple 2D viewpoints. They have incorporated on-screen gestures for fundamental operations. The system was easily mastered and garnered positive feedback from professional users. A similar method was adopted by Ma et al.~\cite{ma20133d} and Xin et al.~\cite{xinNapkinSketchHandheld2008a} in their research. 

Another approach to 3D sketching/modeling allows users to draw a 3D sketch using curves. Subsequently, closed areas are either immediately populated or filled after a certain period with a surface mesh. Grimm et al.\cite{grimmJustDrawIt2012} pioneered this domain, enabling users to transition from a 2D sketch to a 3D curve sketch, followed by patch filling. However, the conversion from 2D to 3D sketches occasionally lacks consistency. Researchers have gravitated towards direct 3D sketching in VR environments as a solution. Proposals for systems specifically designed for VR head-mounted displays (HMDs) have been put forth by De Paoli et al.\cite{depaoliSecondSkinSketchbasedConstruction2015}, Stanko et al.\cite{stankoSmoothInterpolationCurve2016}, and Yu et al.\cite{yuCASSIECurveSurface2021, yuPiecewisesmoothSurfaceFitting2022}. Additionally, a variant of 3D curve sketching exists where an elementary structure is laid out using straight lines, and then 3D curves are incorporated to articulate the desired shape. Such methodologies have been suggested by Schmidt et al.\cite{schmidtAnalyticDrawing3D2009}, and Yu et al.\cite{yuScaffoldSketchAccurateIndustrial2021}.

Research into drawing within a VR environment experienced a resurgence following the introduction of VR HMD hardware,
Arora et al.\cite{aroraExperimentalEvaluationSketching2017} developed a sketching system in VR utilizing the HTC Vive, which allowed artists and designers to design freely in an unconfined space. However, this rudimentary system confronted challenges related to the absence of a physical surface, orientation, and planarity. To facilitate sketching, the researchers employed controllers and 3D pens. To address the concern of the missing physical surface, Elasyed et al.\cite{elsayedVRSketchPenUnconstrainedHaptic2020} incorporated a haptic device within the stylus, which alerts users when they are interacting with the virtual surface. To tackle the planarity issue, Arora et al.\cite{aroraMidAirDrawingCurves2021a} introduced an approach that enabled drawing on any given surface. Meanwhile, Jiang et al.\cite{jiangHandPainter3DSketching2021} presented a sketching technique using a VR tracker to monitor hand positions within the VR environment. They also employed a stylus integrated with optical trackers to enhance positional accuracy.

Users interested in sketching sometimes opt for a virtual 3D canvas. Keefe et al.\cite{keefeCavePaintingFullyImmersive2001a} presented a method that allows users to craft 3D art utilizing 2D brush strokes. Meanwhile, Zeleznik et al.\cite{zeleznikSKETCHInterfaceSketching2006} designed a 3D interface equipped with functionalities such as panning, dragging, rotating, and focusing. Schmid et al.\cite{schmidOverCoatImplicitCanvas2011} and Li et al.\cite{liSweepCanvasSketchbased3D2017} advanced similar systems to construct 3D scenes from RGB-D images that are further refined with sketches. Additionally, Arora et al.\cite{aroraSymbiosisSketchCombining2D2018}, and Drey et al.\cite{dreyVRSketchInExploringDesign2020} incorporated virtual reality into these systems in their respective studies.

\subsubsection{Generating 3D shapes from 3D sketches}
Another approach to 3D modeling exists, which allows users to draw a sketch in 3D and then generate a corresponding 3D model. We have previously delved into the work titled ‘Teddy’ by Igarashi et al.~\cite{igarashiTeddySketchingInterface2006a}. In that method, 2D sketches are extrapolated to form 3D meshes. Another notable instance is FiberMesh~\cite{nealenFiberMeshDesigningFreeform2007}, presented by Nealen et al., a system for designing free-form surfaces with a collection of 3D curves. In this system, users initiate a rudimentary 3D model and then add 3D curves on the model for modification. Andre et al.~\cite{andreSingleviewSketchBased2011} proposed an innovative system capable of producing complex objects from 3D strokes drawn from a unique viewpoint. Their methodology deconstructs each object into smaller primitives defined by construction lines for object generation. Some researchers have also worked on 3D model retrieval using convolutional neural networks from 3D sketches. These research works focus on retrieving 3D models that match the input sketches from the user. Depending on the method, they may return a single object~\cite{li3DSketchBased3D2015, ye3DSketchbased3D2016, lun3DShapeReconstruction2017a}, or multiple models that align with the initial sketch~\cite{wangSketchBased3DShape2015, giunchi3DSketchingInteractive2018}.

\subsubsection{Model guided 3D sketching}
Another variant of 3D sketch-based modeling involves the creation of an object within a 3D space, with guidance from a predetermined shape. Tsang et al.\cite{tsang2004suggestive} introduced an interface where a user can sketch within a 3D space and use a 2D image as a guide. The strokes input by users also triggered geometric suggestions, reducing the requirement to draw all parts of the new model. Similarly, Zheng et al.\cite{zhengSmartCanvasContextinferredInterpretation2016} developed a system tailored for architectural or design-oriented sketches. Here, users render a 2D sketch derived from an image of an architecture, after which the system converts the 2D sketch into an intricate 3D model based on some primitive shapes. Jackson et al.\cite{jacksonLiftOffUsingReference2016} proposed an analogous system specifically for transmuting 2D sketches into 3D models within a VR environment. In a distinct approach, Yue et al.\cite{yueWireDraw3DWire2017a} introduced a technique for crafting a 3D sketch utilizing a specialized 3D extruder pen, where users receive guidance from a virtual model projected through a VR headset. Xu et al.~\cite{xuModelGuided3DSketching2019} presented a novel interface specifically for 3D model-guided sketching on 3D planes. This methodology strategically leverages the geometry of an underlying 3D model to infer 3D planes on which 2D strokes were drawn and around the mentioned 3D model.

\section{Interaction Tools}\label{sec:inter}
For free-form modeling, the interaction tools take a significant portion of the user experience. Our study of the literature reveals that various modeling techniques highlight the impact of different interaction tools and methods on user attention and expertise. We also discovered that certain devices are preferred for sculpting while others are favored for sketching. Additionally, there are tools such as pen-based tools, which are considered standard for both modeling techniques, like pens for sketching. Furthermore, we came across mentions of unique interactive tools such as 3D positional trackers, custom stylus, etc. In this section, we will explore these different interaction tools and examine how they provide users with advantages in free-form shape modeling compared to other devices.

\subsection{Gestures-based tools}
Gestures, especially hand gestures, are among the most common forms of interaction. For sculpting as they provide the users a natural experience to interact with the shapes or objects \cite{kimTangible3DHand2011, mcdonnellVirtualClayRealtime2001a, wessonEvaluatingOrganic3D2013, gaoDigiClayInteractiveInstallation2018a}. There are also approaches for using hand gestures as the interactive medium for drawing \cite{schkolneSurfaceDrawingCreating2001a, kimSketchingWithHands3DSketching2016, amoresHoloARtPaintingHolograms2017a, dudleyBareHanded3DDrawing2018a, cohen3DVirtualSketching2019a, keefeCavePaintingFullyImmersive2001a}. Nowadays, depth cameras such as Kinect and Leap Motion can capture hand gestures accurately. However, back in 2004, Tosas et al. \cite{tosasVirtualTouchScreen2004a} proposed a method that employed a camera, a black screen, and a virtual grid to capture various gestures. It was not competent enough to capture moving hand gestures, but many other researchers proposed methods based on this for capturing hand gestures later~\cite{tosas2007virtual, gallotti2011v}.

McDonnell et al. \cite{mcdonnellVirtualClayRealtime2001a} proposed a modeling framework and haptic feedback based on physics-based techniques. They used virtual clay as their sculpting interface, with several tools like finger tracking and 3D pens for control. Similar principles were applied by Kim et al. \cite{kimTangible3DHand2011} and Gao et al. \cite{gaoDigiClayInteractiveInstallation2018a} where they both used hand gestures in the air to sculpt objects virtually. Although their methods were similar but, hand gesture recognition differed. Kim et al. used pictures of hands taken similarly to Tosas et al. \cite{tosasVirtualTouchScreen2004a}, and Gao et al. used Kinect to recognize the gestures. Wesson et al. \cite{wessonEvaluatingOrganic3D2013} proposed a sculpting and evaluation technique using Kinect to mimic the interaction of physical sculpting. Pihuit at al. \cite{pihuitHandsVirtualClay2008} proposed a different method where, instead of capturing hand gestures, they used a pressure-sensitive device to mimic the hand pressure for sculpting. In this way, users were also getting haptic feedback on their input. Later, Jacobs et al.~\cite{jacobsSoftHandModel2011} used a soft hand model for virtual manipulation.

For 3D sketching, hand gestures are not always a preferred tool, as they lack haptic feedback and may not produce accurate sketches without augmented guidance. However, there are few papers~\cite{schkolneSurfaceDrawingCreating2001a, amoresHoloARtPaintingHolograms2017a, dudleyBareHanded3DDrawing2018a, kwanMobi3DSketch3DSketching2019} on sketching with hand gestures. Schkolne et al. \cite{schkolneSurfaceDrawingCreating2001a} proposed a method for drawing surfaces with several interaction tools, including hand gestures captured by an electronic glove. Drawing surfaces may not require as much accuracy as drawing curves or lines, making the hand a suitable choice, but, for some regions, users may prefer specialized tools over hands~\cite{schkolneSurfaceDrawingCreating2001a}. Kim et al. \cite{kimAgile3DSketching2018} proposed drawing a scaffold of an object where users do not have to be precise, using the object as a guideline. There are some proposals~\cite{amoresHoloARtPaintingHolograms2017a, dudleyBareHanded3DDrawing2018a, kimSketchingWithHands3DSketching2016} where hands, especially finger motions, were used as the tool for drawing, where the user uses fingers to draw in the air and has a guideline to lead the stroke. There was also some research by Kwan et al.~\cite{kwanMobi3DSketch3DSketching2019} and Dudley et al.~\cite{dudleyBareHanded3DDrawing2018a} on 3D sketching using hands in an AR environment.

\subsection{Pen-based tools}
Pen or stylus-based tools are another popular standard for interaction with sketches and sculpting. This type of tool provides precise control to the user. With pen-like tools, users can interact with the object accurately, as it simulates the natural motion of sketching. In our study, we found that different types of pen-based tools have been used in past research. In our study, we learned that pen-based tools are mainly used for sketching and surface drawing, but there exists a small body of research on sculpting with a pen-based tool \cite{galyeanSculptingInteractiveVolumetric1991a, zoranFreeDFreehandDigital2013b}. 

The most common type of pen-based tool used for sketching is the stylus for drawing pads. This method is widely used, even for 3D sketching, where users sketch on a 2D tablet screen while working in 3D. Using a drawing tablet provides users with a specific projection and haptic feedback. Keefe et al. \cite{keefeCavePaintingFullyImmersive2001a, keefeDrawingAirInput2007a}, Bae et al. \cite{baeILoveSketchAsnaturalaspossibleSketching2008a}, and Xin et al. \cite{xinNapkinSketchHandheld2008a} have explained this technique in their research, which demonstrates that users can represent and interact with 3D sketches on a 2D screen \cite{baeEverybodyLovesSketch3DSketching2009a, ma20133d}. Paczkowski et al. \cite{paczkowskiInsituSketchingArchitectural2011}, Kazi et al. \cite{kaziDreamSketchEarlyStage2017}, and Arora et al. \cite{aroraSymbiosisSketchCombining2D2018} have extended this idea to a next level by adding drawing in AR environment. In another method, Kim et al. \cite{kimSketchingWithHands3DSketching2016} have proposed that users start the sketch with an impression of the hand and then add details using a pen.

Custom-made 3D stylus with 3D tracking sensors is another notable pen-based tool that has gained prominence and is primarily designed for 3D sketching purposes. Numerous articles~\cite{kamuroHapticEditorCreation2012, jacksonLiftOffUsingReference2016, yueWireDraw3DWire2017a, hoffmann2023thermalpen} delve into the functionalities and applications of this tool. Its primary usage revolves around sketching and drawing within VR environments. In some cases, the pen's enclosure incorporates multiple 3D tracking sensors, as proposed by Kamuro et al. \cite{kamuroHapticEditorCreation2012} and Jackson et al. \cite{jacksonLiftOffUsingReference2016}. Furthermore, Yue et al. \cite{yueWireDraw3DWire2017a} presented a compelling proposal for a customized 3D extruder pen equipped with a 3D positioning sensor. Additionally, the method introduced by Arora et al. \cite{aroraSymbiosisSketchCombining2D2018} offers a 3D sketching option, wherein the drawing tool operates based on a similar conceptual framework. The continuous exploration and innovation in the field of pen-based tools for three-dimensional artistic expression are reflected in these advances.

\subsection{3D Controller}
In our study, we found that in some methods~\cite{luo3DVRSketch3D2020, dashtiPotteryVRVirtualReality2022, gaoRealPotImmersiveVirtual2019, xu2023gesturesurface}, VR controllers were used as an interaction tool. This specific type of shape modeling method was designed for the VR environment so that the controllers gave native support. There are also a few researchers who created a custom tracker for sketching or sculpting. Eroglu et al. \cite{erogluFluidSketchingImmersive2018},  Dong et al. \cite{dongRealTimeReTexturedGeometry2018}, and Jiang et al. \cite{jiangHandPainter3DSketching2021} designed their custom controller based on the 3D positional trackers. There were also methods proposed by Giunchi et al. \cite{giunchi3DSketchingInteractive2018} and Garcia et al. \cite{garciaSpatialMotionDoodles2019} where they used the native VR controllers for interaction.

\subsection{Other Tools}
In our study, we found a few more interaction tools that are exclusive to the aforementioned ones. There were some studies by Wang et al. \cite{wangSketchUpVirtualWorld2008}, and Hagbi et al. \cite{hagbiInPlaceSketchingContent2010}, where they designed sketching systems, which used a mouse as their interaction tool. Although they used similar interaction tools, Wang et al.~\cite{wangSketchUpVirtualWorld2008} proposed a method to construct 3D shapes from 2D sketches in the virtual world, whereas Hagbi et al.~\cite{hagbiInPlaceSketchingContent2010} worked on drawing in AR. Another technique we found involves cutting \cite{marnerAugmentedFoamSculpting2010a} or sculpting \cite{zoranFreeDFreehandDigital2013b} a piece of foam or creating paper-crafts \cite{ohPEP3DPrinted2018a} to form a shape and fabricate it into a CAD model.

\section{AI in modeling}\label{sec:AI} 
With newer research, different AI techniques are now integrated into the space of shape modeling. The AI modeling techniques we discovered in our research can be categorized into suggestive modeling and AI-based shape generation. In the former technique, the system suggests future edits, and in the latter, the system generates a model based on specified constraints. Both approaches offer unique advantages and potential applications in various domains.

\subsection{Suggestive Modeling}
\begin{figure*}[!ht]
     \centering
     \includegraphics[width=.95\linewidth]{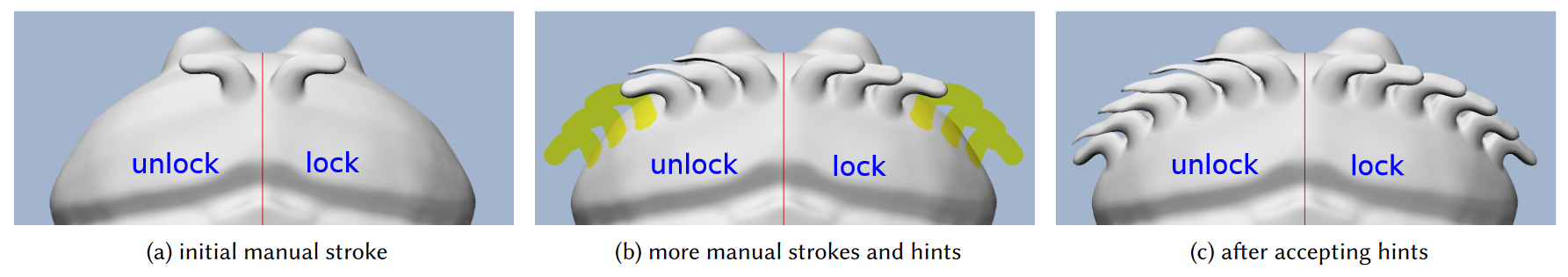}  
     \caption{An example of automated sculpting. New strokes are automatically added based on the previous stroke. ~\cite{pengAutocomplete3DSculpting2018a}.}
     \label{fig: auto}
\end{figure*}

It is widely acknowledged that creating drawings or sculptures in the air with gestures without haptic feedback is challenging, as users do not experience any tactile feedback~\cite{kamuroHapticEditorCreation2012}. Additionally, any kind of 3D modeling can be difficult for a user with limited or no experience. To address these issues, researchers have developed suggestive modeling. This technique uses machine learning or AI to suggest users the object that they aim to create. These suggestions are generated through the analysis of the user's past modeling efforts or by trying to recognize the shape that they are attempting to create. The system not only gives suggestions for future sculpting actions, but it can also correct any mistakes as well. Iaruss et al. \cite{iarussiDrawingAssistantAutomated2013} and Peng et al. \cite{pengAutocompleteAnimatedSculpting2020a} proposed this type of approach, using deep learning to recognize previous shapes and provide suggestions for improvement. Peng et al.~\cite{pengAutocomplete3DSculpting2018a} also conducted research on automating the modeling process by mirroring or predicting the user's actions (as shown in Figure~\ref{fig: auto}). For 3D sketching, there are methods for model-guided sketching described by Xu et al.~\cite{xuModelGuided3DSketching2019}, and El Sayed et al.~\cite{elsayedVRSketchPenUnconstrainedHaptic2020} that guide users for the upcoming strokes.  Guillard et al. \cite{guillardSketch2MeshReconstructingEditing2021a} and Zhang et al. \cite{zhangSketch2ModelViewAware3D2021} have also presented frameworks that take a sketch input and generate a 3D mesh or model to represent the shape, allowing the user to perform further edits.

\subsection{Shape Generation using AI}
Sculpting or sketching shapes can be challenging for novices. To address this, researchers have explored using artificial intelligence to assist users in modeling figures. Typically, an input is given in the form of a sketch or image, and a more precise 3D shape is generated, represented in mesh~\cite{gkioxari2019mesh, kundu20183d}, voxel~\cite{gkioxari2019mesh}, or implicit~\cite{park2019deepsdf, chabra2020deep, genova2020local, tzathas20233d} data structures. From either single-view or multi-view images, users can produce a 3D structure. On occasions, these 3D models are editable~\cite{kundu20183d, tzathas20233d}. However, editing capabilities are typically restricted to one model at a time.

In 2020, Mildenhall et al.~\cite{mildenhall2021nerf} introduced the groundbreaking concept of the Neural Radiance Field (NeRF). This innovative framework offered a novel approach to 3D scene generation. A distinct advantage of the NeRF is its ability to bypass the requirement of possessing data corresponding to each individual frame. Concurrently, advancements in AI-driven image generation techniques have facilitated the creation of novel views from singular image via the application of the diffusion model~\cite{rombach2022high}. An intriguing convergence of technologies arises when one combines the capabilities of the diffusion model with the foundational principles of the NeRF. By integrating the diffusion model with NeRF, it becomes possible to instantiate an object solely from an image or a textual prompt~\cite{poole2022dreamfusion}. Subsequent to their inception, both techniques went through refinements to enhance the fidelity of the output~\cite{barron2021mip, barron2022mip, pumarola2021d} and to speed up~\cite{yu2021plenoxels, pupezescu2022instant, li2022nerfacc} the scene generation process. Beyond these, researchers have devised even more intricate methodologies for converting images or textual prompts into 3D representations~\cite{melas2023realfusion, lin2023magic3d} have been developed. In a noteworthy contribution to this field was NeRFShop by Jambon et al.~\cite{jambon2023nerfshop}, who proposed a methodology allowing users to interact directly with the generated scenes. This not only streamlines the creation of a desired object but also provides users the ability to modify it post-generation. Mikhaeli et al.~\cite{mikaeili2023sked} proposed SKED, where a NeRF model is edited using sketches and text prompts.

\section{Collaborative Editing}\label{sec:collab} 
\begin{figure}[!ht]
     \centering
     \includegraphics[width=.95\linewidth]{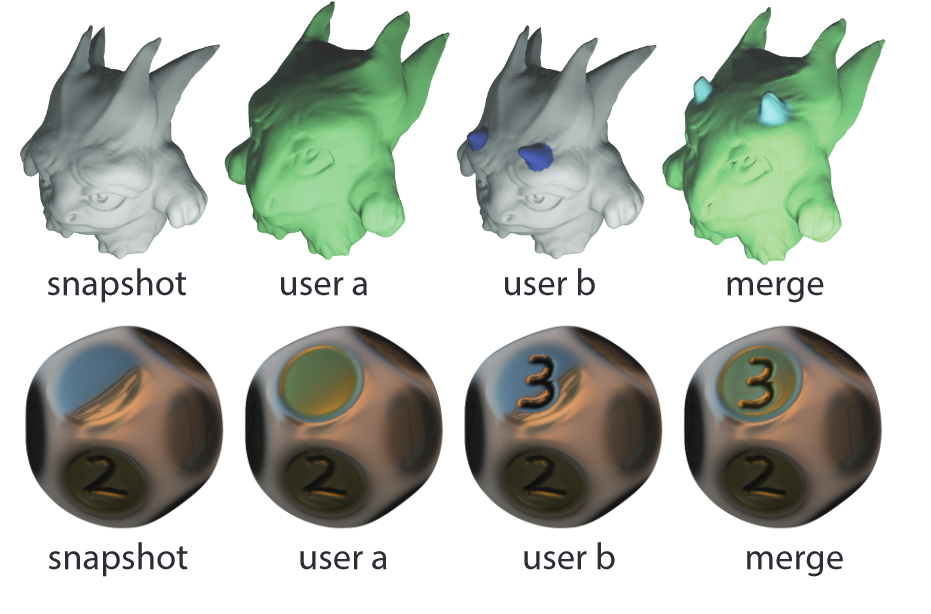}  
     \caption{Collaborative modeling between two users \cite{calabreseCSculptSystemCollaborative2016b}.}
     \label{fig: collab}
\end{figure}

It is common for users and designers to want to present their models or drawings to others for showcasing or review purposes. Collaborative modeling enables multiple individuals to work on the same object simultaneously. It presents its own challenges, such as conflicting edits to the same section of the model, and latency issues when designers are working remotely. Hernández et al. \cite{hernandezWeSketch3DReal2011} proposing an innovative, easy-to-use, and optimized interface, Denning et al. \cite{denningMeshGitDiffingMerging2013b} presented a "diff and merge" method where multiple users can work on separate parts of the model and merge their work into the main model later. Calabrese et al. \cite{calabreseCSculptSystemCollaborative2016b} (see Figure~\ref{fig: collab}) proposed a unique collaboration technique where changes made by two or more users can be merged by taking either the average, a cumulative or weighted sum of the changes. Hoppe et al. \cite{hoppeShiShaEnablingShared2021a} designed a system that allows users to work both locally and remotely simultaneously.

\section{Discussion}
We have observed that free-form shape modeling has garnered significant attention due to the advancement of affordable XR tracking and display technologies. This has subsequently captivated the interest of numerous researchers and practitioners in the domain. Earlier, we mentioned that we aim to address several questions that emerged during our discussion on methodology (see section ~\ref{sec:method}). After conducting a thorough review, we have been able to provide answers to these questions.

\begin{figure*}[!htp]
    \centering
    \begin{tabular}{cc}
        \includegraphics[width=0.4\textwidth]{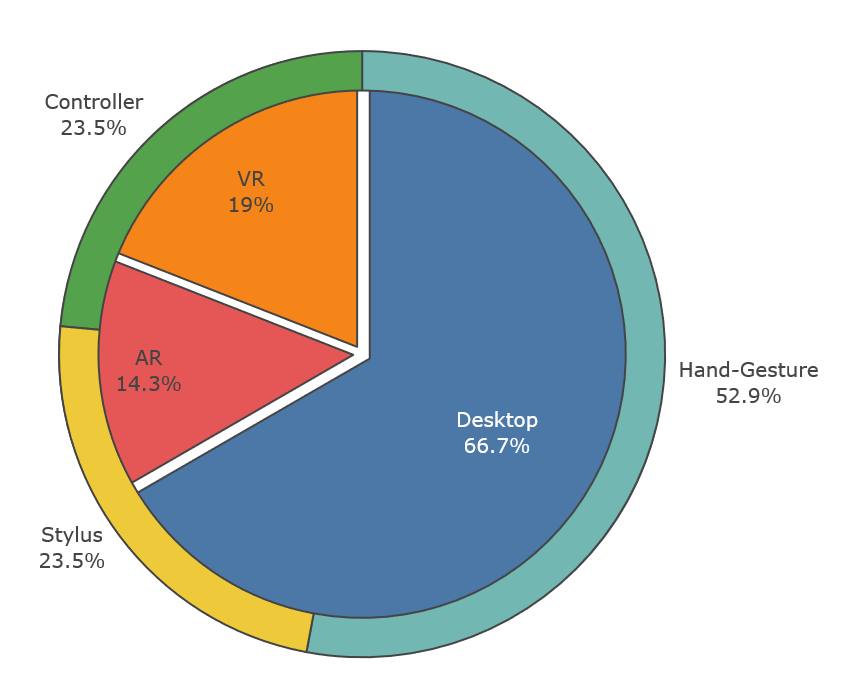} & \includegraphics[width=0.4\textwidth]{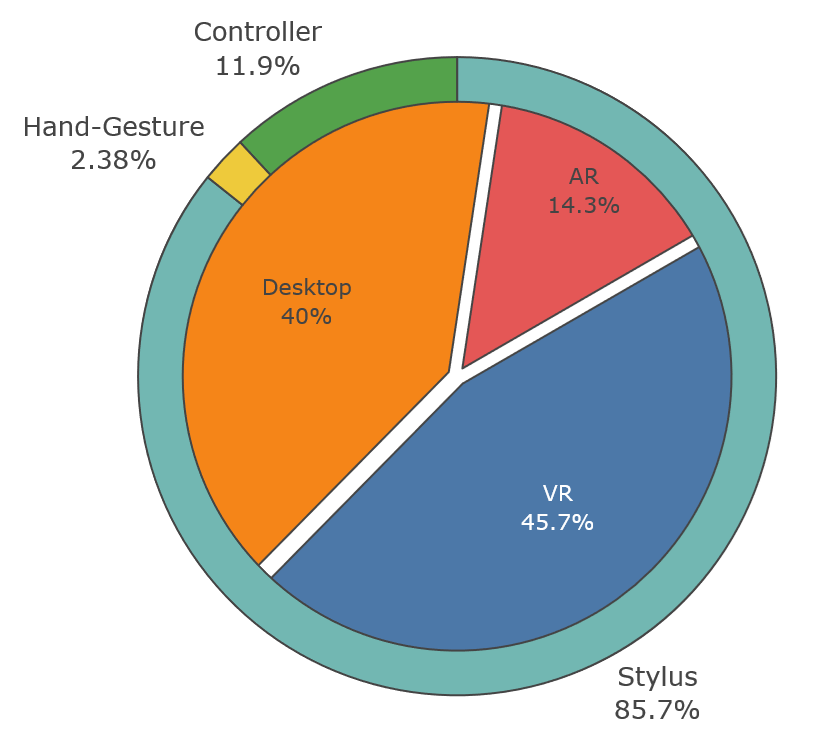} \\
        (a) Sculpting & (b) Sketching
    \end{tabular}
    \caption{The trend for research in different environments and interaction tool for sculpting and sketching methods.}
    \label{fig: measure}
\end{figure*}

\subsection{Preferred Techniques for modeling and their methods} Our findings suggest that 3D sketching is a preferred modeling technique compared to virtual sculpting. Based on the articles analyzed for this study, it can also be concluded that researchers exhibit a preference for the 3D sketching method. This conclusion is derived from the abundant and diverse research available on sketching techniques as opposed to that on sculpting techniques. Furthermore, this preference may arise because users can start from scratch with 3D sketching and gradually progress, while sculpting may be unfamiliar or challenging to receive adequate feedback on. In contrast, various methods have been developed for sketching with a stylus and tablet, some of which replicate the pen and paper experience through vibrations.

During our research, we encountered diverse methodologies associated with virtual sculpting and 3D sketching. Although we attempted to categorize the majority of these methods, it remains plausible that certain techniques may have been overlooked. In the realm of sculpting, we identified notable techniques such as virtual clay, wherein users modify a filled shape. Conversely, in surface modeling techniques, users adjust the surface of a hollow form. Another identified method is sculpting with a physical proxy, in which the user replicates editing actions on a tangible object. The material is segregated into layers to incorporate intricate details in multi-layer sculpting. Similarly, our study on 3D sketching unveiled methodologies like generating 3D shapes either from 2D or 3D sketches, sketching curves within 3D spaces, and employing a projected model as a guide during 3D sketching. Additionally, our study touched upon AI shape modeling techniques utilizing radiance fields. In Figure ~\ref{fig: measure}, we have shown the research trend for sculpting and sketching. It is a metric that shows the involvement of different environment and interaction tools.

\subsection{Research Gaps} Upon scrutinizing our study, we identified several research gaps. However, some of them seem to recur, appearing with notable frequency across our observations and analyses. Although using VR or MR headsets in 3D shape modeling undeniably offers discernible benefits, they fall short in replicating the tactile feedback experienced in tangible, real-world scenarios. Additionally, the occasional latency experienced within virtual environments can considerably diminish immersive user experience. This can be attributed to the inherent computational constraints of contemporary hardware. Notably, there appears to be significant potential for advancements in AI shape modeling, an area presently constrained by the limits of current computational capabilities.

\subsection{Preferred Hardware} Our analysis compellingly indicates that, for 3D sketching in any environment, users exhibit a strong inclination towards tools that closely emulate the tactile experience of drawing with a pen, capturing the intuitive essence of the traditional pen-and-paper approach. On the other hand, gestures emerged as the preferred tools for sculpting. They offer the dual advantage of encompassing larger regions for shaping while still affording the user the precision necessary for detailed work. Yet, it's undeniable that auxiliary tools are crucial for ultra-fine operations. It's imperative that future research robustly addresses the palpable challenges of delivering apt feedback for both sculpting and sketching in a virtual 3D milieu. Additionally, the seamless incorporation of AI and collaborative infrastructures into these methods beckons more in-depth exploration. The intricacies of managing simultaneous alterations by multiple contributors during cooperative modeling persist as a monumental hurdle.

In recent years, there has been an increase in the attention given to the field of AI model generation. This surge in interest reflects the potential and importance of artificial intelligence in shaping the technological landscape. However, as with many emerging technologies, it is not without its challenges. Presently, the algorithms that drive the creation of AI models tend to operate at a pace that some may find sub-optimal. This slow rate of generation can hinder rapid progress, especially in applications that demand quick turnarounds. Moreover, the AI models' qualitative aspects pose a significant concern. For VR, the experience is largely determined by the fidelity and quality of the underlying models. The present AI models, unfortunately, do not always meet the high standards required to create a truly immersive VR experience.

Another aspect of the AI model generation process that currently lags behind is the subsequent editing and modification phase. The tools and methodologies currently available for this purpose exhibit a lack of intuitive design, even in the desktop environments. Presently, many of these editing methods are reliant on text-based prompts, which introduces another layer of complexity. The use of text prompts for model editing is filled with challenges. For one, with the text prompt, users lose artistic control. When relying on textual instructions, an inherent ambiguity can limit the precision with which one can modify the AI model. Moreover, a seemingly minor alteration or update to a prompt can sometimes trigger a dramatic change in the resultant model. In applications where continuity is paramount, such as in the creation of digital assets, this unpredictability can be particularly problematic.

However, researchers are diligently working on developing high-quality and rapid 3D representations that can be modified even after their initial generation. Presently, for images, we possess the capability to modify them through techniques like in-painting, dragging, and text prompts. Drawing inspiration from these advancements, researchers are now striving to adapt and implement similar modifications for 3D-generated models. It is anticipated that in the near future, such manipulations of 3D models will become commonplace and widely accessible.

\section{Conclusion}\label{sec:conclusion}
In this survey, we presented state-of-the-art free-form shape modeling in extended reality. We began by examining the desktop environments, then expanded our survey to include augmented and virtual reality as mediums for shape modeling. Our survey compares techniques of free-form shape modeling, such as sculpting and 3D sketching, and discusses the research on both algorithmic and interactive aspects. Based on the collected set of publications that encompasses all relevant techniques, we derived a taxonomy of free-form shape modeling consisting of three main categories: the field of contribution, either on the algorithm side or the interaction side; the working environment, such as a desktop, AR, or VR; and the type of interaction method used, such as gestures, controllers or pen-like tools, even though gaps in the taxonomy hint at the areas where work has been done, as well as hint at opportunities to fill gaps in existing research literature. We also identified significant challenges that could guide future research. With the contributions mentioned, our aspiration is for an enhanced comprehension of free-form shape modeling.

\bibliographystyle{unsrt} 
\bibliography{egbibsample}

\end{document}